\documentstyle [12pt] {article}
\newcommand{\cs}[3]{{{#3} \brace {#1 #2}}}
\begin{document}
\newcommand{\h}[1]{\mathop{\lambda}\limits_{#1}\ \!\!\!}
\newcommand{\pder}[2]{\frac{\partial{#1}}{\partial{#2}}}
\newcommand{\tder}[2]{\frac{d{#1}}{d{#2}}}
\begin{center}
{\huge Quantum Features of Non-Symmetric Geometries }
\end{center}
\begin{center}
${{\bf M.I.Wanas}~^1{\bf and ~M.E.Kahil}}$
\footnote[1]{Astronomy  Department, Faculty of Science, Cairo University,  Giza, Egypt.\\
~~~e-mail: wanas@frcu.eun.eg} 
\end{center}
 
\begin{center}
\section*{\bf Abstract}
\end{center}
Paths in an appropriate geometry are usually used as trajectories of test
particles in geometric theories of gravity. 
It is shown that non-symmetric geometries possess some interesting quantum
features. Without carrying out any quantization schemes, paths in such
geometries are naturally quantized. Two different non-symmetric geometries
are examined for these features. It is proved that, whatever the non-symmetric
geometry is, we always get the same quantum features.
It is shown that these features appear only in the pure torsion term (the 
anti-symmetric part of the affine connection) of the path equations. The 
vanishing of the torsion 
leads to the disappearance of these features, regardless of the symmetric part
of the connection. It is suggested that, in order to be consistent with the 
results of experiments and observations, torsion term in path 
equations should be parametrized using an appropriate parameter.\\
\begin{center} 
\bf{KEY WORDS: Quantum - Non-Symmetric Geometries - Torsion}
\end{center}

\section*{\bf 1. Introduction}

It is well
known that the theoretical description of gravitational interaction is successful in the framework of geometrization. In this framework trajectories of test particles, in a background 
gravitational field, are represented by path equations in the geometry used. For example, in case of General Relativity, written in Riemannian geometry, the equations of motion are geodesic 
(and/or null-geodesic) equations. Also, it is known that quantization of gravitational interactions is not, so far, successful. The reason may be rooted in the conventional schemes used for 
tackling the problem of quantization.

Keeping these points in mind, it is of interest to examine the properties  of any appropriate geometry, especially 
those concerning path equations, to look for quantum features, if any. As will appear in due course Riemannian geometry is not a suitable candidate for our purpose. The natural modification {( or generalization )} of this geometry is the non-symmetric geometry. We mean by a non-symmetric geometry, a geometry in which one can define a non-symmetric affine connection, i.e. a geometry admitting torsion. This class of geometry includes the absolute parallelism (AP)geometry 
[1]. It includes also the non-symmetric geometry used by Einstein in his last trial for constructing a unified field theory [2].

In a previous work [3], it has been shown that new path equations in the absolute parallelism (AP)-geometry could be derived using a certain variational approach  given by Bazanski [4] . These equations can be written in the following form
\footnote{We will use the parentheses {( )} for symmetrization and
  brackets {[ ]} for antisymmetrization.},

\begin{equation}
{{\frac{dJ^\mu}{dS^-}} + \cs{\nu}{\sigma}{\mu}\ J^\nu J^\sigma = 0},
\end{equation}
\begin{equation}
{{\frac{dW^\mu}{dS^o}} + \cs{\nu}{\sigma}{\mu}\ W^\nu W^\sigma = 
- {\frac{1}{2}}\Lambda_{(\nu \sigma)} . ^\mu~~ W^\nu W^\sigma},
\end{equation}    
\begin{equation}
{{\frac{dV^\mu}{dS^+}} + \cs{\nu}{\sigma}{\mu}\ V^\nu V^\sigma = 
- \Lambda_{(\nu \sigma)} . ^\mu~~ V^\nu V^\sigma} ,
\end{equation}    
where $J^\mu$ ,$W^\mu$ and $V^\mu$ are the tangents to the corresponding curves whose parameters are $S^-$,$S^0$, and $S^+$ respectively , $\cs{\nu}{\sigma}{\mu}$ is the Christoffel
symbol of the second kind and $\Lambda^\alpha _{~\mu \nu}$ is the torsion of the AP-geometry defined by  
\begin{equation}
\Lambda^\alpha_{~ \mu \nu} {\stackrel{def.}{=}}~ \Gamma^\alpha_{. \mu \nu} - 
\Gamma^\alpha_{. \nu \mu},  
\end{equation}
where $\Gamma^\alpha_{. \nu \mu}$ is the non -symmetric affine connection defined
as a condition for AP,viz
\begin{equation}
\h{i}^\alpha_{+~|\mu} = 0 ,
\end{equation}
where $\h{i}^{\alpha}$ are the
tetrad vectors defining completely the structure of AP-space in 4-dimensions. These equations 
are considered to represent generalization of geodesic equations of the Riemannian 
geometry.

Two interesting notes could be reported regarding these equations:\\
(i) Recalling that the unique symmetric affine connection of Riemannian geometry
gives rise to the geodesic path, it is shown that in the present case
the path equations (1),(2) and (3) are derived using  four affine connections 
defined in the AP-geometry, $\Gamma^\alpha_{. \mu \nu}$,$\hat\Gamma^\alpha_{. \mu \nu} (= \Gamma^\alpha_{. \nu \mu})$, $\Gamma^\alpha_{. (\mu \nu)}$ and $\cs{\nu}{\sigma}{\mu}$ .\\
(ii) The numerical coefficients of torsion term in equations (1),(2) and (3) are 0,$\frac{1}{2}$
and 1 respectively, i.e. the coefficient jumps by a step of one half from one equation to the 
next.\\

Now it is of interest to ask the following questions :\\
 (a) If other affine connections are defined in the AP-geometry, what are the corresponding 
 
 path equations, if any ? \\
 (b) Does the value of numerical step (one half in the above set of path equations) 

 differ by changing the affine connections?\\
 (c) If we have another type of non -symmetric geometry, what are the resulting equations?

Does the numerical step of torsion term  remain unchanged ?\\
 The present work is an attempt to answer the above questions. In section 2, we examine the consequences of using affine connections different from those used in the previous work.
In section 3 we review briefly the bases of another type of non-symmetric  geometry. 
In section 4 we give the path equations for the geometry given in section 3 
 . We discuss and conclude the work in section 5.\\ 
\section*{\bf 2. Quantum Features of the AP-Geometry}

 In addition to the affine connections mentioned in the previous section, we 
can define two further connections, ${\Omega}^\alpha_{.\mu \nu}$ and
 ${\hat\Omega}^\alpha_{.\mu \nu}$(=${\Omega}^\alpha_{.\nu \mu}$), as 
\begin{equation}
{\Omega}^\alpha _{.~\mu \nu} {\stackrel{def.}{=}}~  \cs{\mu}{\nu}{\alpha}\ + \Lambda^\alpha_{.~\mu \nu}.
\end{equation}
Using these two connections one can define the following absolute covariant derivatives: 
\begin{equation} 
A^{\mu}_{+||~ \nu} = A^{\mu}_{~, \nu} + A^{\alpha}~{\Omega}^{\mu}_{.~\alpha \nu},
\end{equation}
\begin{equation}
A^{\mu}_{-||~ \nu} = A^{\mu}_{~, \nu} + A^{\alpha}~{\hat\Omega}^{\mu}_{.~\alpha \nu},
\end{equation}
\begin{equation}
A^{\mu}_{0||~ \nu} = A^{\mu}_{~, \nu} + A^{\alpha}~{\Omega}^{\mu} _{.~(\alpha \nu)}=A^{\mu}_{~; \nu} ,
\end{equation}
 where $A^\alpha$ is an arbitrary vector and (;) denotes covariant differentiation using Christoffel symbol. The derivatives (7),(8) and (9) are related to the parameter derivatives by the relations
\begin{equation}
\frac{DA^\mu}{D \hat S^{-}} = A^{\mu}_{-||~ \alpha} \hat J^\alpha,
\end{equation}
\begin{equation}
\frac{DA^\mu}{D \hat S^{0}} = A^{\mu}_{0||~ \alpha} \hat W^\alpha,
\end{equation} 
\begin{equation}
\frac{DA^\mu}{D \hat S^{+}} = A^{\mu}_{+||~ \alpha} \hat V^\alpha,
\end{equation}  
where $\hat S^{-}$,$\hat S^{0}$and $\hat S^{+}$ are the parameters varying along the corresponding curves whose tangents are ${\hat J}^\alpha$,$\hat W^\alpha$ and ${\hat V}^\alpha$ respectively.

Applying the Bazanski approach [4] to the Lagrangians,
\begin{equation}
L^{-} = \h{i}_{\alpha}~ \h{i}_{\beta}~ \hat J^{\alpha}~ {\frac{{\bf{D}}{\eta}^{\beta}}{{
\bf{D}} \hat S^{-}}} ,  
\end{equation}
\begin{equation}
L^{0} = \h{i}_{\alpha}~ \h{i}_{\beta}~ \hat W^{\alpha}~ {\frac{{\bf{D}}{\zeta}^{\beta}}{{
\bf{D}} \hat S^{0}}} ,     
\end{equation}
\begin{equation}
L^{+} = \h{i}_{\alpha}~ \h{i}_{\beta}~ \hat V^{\alpha}~ {\frac{{\bf{D}}{\xi}^{\beta}}{{
\bf{D}} \hat S^{+}}} , 
\end{equation} 
where $\eta^\beta$, $\zeta^\beta$ and $\xi^\beta$ are deviation vectors, we get the following set of path equations respectively:
 \begin{equation}
{{\frac{d\hat J^\mu}{d\hat S^-}} + \cs{\nu}{\sigma}{\mu}\ \hat J^\nu \hat J^\sigma = 0},
\end{equation}
\begin{equation}
\frac{d\hat W^\mu}{d\hat S^o} + \cs{\nu}{\sigma}{\mu}\ \hat W^{\nu} \hat W^{\sigma} = - 
\frac{1}{2} \Lambda_{(\nu \sigma)}^{..~~\mu}~~ \hat W^{\nu} \hat W^{\sigma} ,
\end{equation}    
\begin{equation}
\frac{d\hat V^\mu}{d\hat S^+} + \cs{\nu}{\sigma}{\mu}\ \hat V^\nu \hat V^\sigma = 
- \Lambda_{(\nu \sigma)}^{..~~\mu}~~ \hat V^{\nu} \hat V^{\sigma} ,
\end{equation}   
which is the same set obtained in the previous work and given in section 1, with the same numerical step characterizing the torsion term.

 Now we will summarize the results obtained in AP-geometry. Let us 
write the set of the path equations, obtained in this geometry, in the following form:
\begin{equation}
{\frac{dB^\mu}{d\hat S}} + a~ \cs{\alpha}{\beta}{\mu}\ B^{\alpha} B^{\beta} = - b~ \Lambda_{(\alpha \beta)}^{~.~.~~\mu}~~B^{\alpha} B^{\beta},
\end{equation}
where $a$, $b$ are the numerical coefficients of the Christoffel symbol term and of the 
torsion term respectively. Thus we can construct the following table.
\begin{center}
 Table I: Numerical Coefficients of The Path Equation in AP-Geometry      \\ 
\vspace{0.5cm}
\begin{tabular}{|c|c|c|} \hline
& & \\
Affine Connection  &Coefficient $a$  &Coefficient $b$   \\
& & \\ \hline
& & \\ 
${\hat\Gamma}^{\alpha}_{.~\mu \nu}$     & 1   & 0  \\
& & \\ \hline
& & \\ 
${\Gamma}^{\alpha}_{.~(\mu\nu)}$      &  1   & $\frac{1}{2}$ \\
& & \\ \hline
& & \\
${\Gamma}^{\alpha}_{.~\mu \nu}$ &1 &1 \\
& & \\ \hline
& & \\ 
${\hat\Omega}^{\alpha}_{.~\mu \nu}$  &1      &0  \\
& & \\ \hline
& & \\
${\Omega}^{\alpha}_{.~(\mu \nu)} =\cs{\mu}{\nu}{\alpha}\   $   & 1 &  $\frac{1}{2}$ \\
& & \\ \hline
& & \\ 
${\Omega}^{\alpha}_{.~\mu \nu}$   & 1    &1 \\  
& & \\ \hline
\end{tabular}
\end{center}            
\section*{\bf 3. Einstein's Non-Symmetric Geometry}

Einstein generalized the Riemannian geometry by droping the symmetry conditions
imposed on the metric tensor and on the affine connection [2].
In this geometry the non-symmetric metric tensor is given by
\begin{equation}
g_{\mu \nu} {\stackrel{def.}{=}}~ h_{\mu \nu} + f_{\mu \nu} , 
\end{equation}  
where $$
        h_{\mu \nu} {\stackrel{def.}{=}}~ \frac{1}{2}( g_{\mu \nu} + g_{\mu \nu})   ,     
$$
$$    
     f_{\mu \nu} {\stackrel{def.}{=}}~ \frac{1}{2}(g_{\mu \nu}  -  g_{\mu \nu})   .
$$
Since the connection of the geometry $U^{\alpha}_{.~\mu \nu}$ is assumed to be
non-symmetric, we can define the following 3-types of covariant derivatives:
\begin{equation} 
A^{\mu}_{+|||~ \nu} {\stackrel{def.}{=}}~ A^{\mu}_{~, \nu} + A^{\alpha}U^\mu _{.\alpha \nu} ,
\end{equation}
\begin{equation}
A^{\mu}_{-|||~ \nu} {\stackrel{def.}{=}}~ A^{\mu}_{~, \nu} + A^{\alpha}U^\mu_{.\nu \alpha} ,
\end{equation}
\begin{equation}
A^{\mu}_{0|||~ \nu} {\stackrel{def.}{=}}~ A^{\mu}_{~, \nu} + A^{\alpha}U^\mu _{.(\alpha \nu)} ,
\end{equation}         
where $A^\mu$ is any arbitrary vector. Now the connection $U^{\alpha}_{.~\mu \nu}$
is defined such that [2],
\begin{equation}
g_{\stackrel{\mu}{+} \stackrel{\nu}{-} |||\sigma} = 0 ,
\end{equation}
\begin{equation}    	 
i.e.~~~ g_{\mu \nu ,\sigma} = g_{\mu \alpha} U^{\alpha}_{.~~\sigma \nu} +
g_{ \alpha \nu} U^{\alpha}_{.~~\mu \sigma}.  
\end{equation} 	 
The non-symmetric connection can be written in the the following form:
\begin{equation}
U^{\alpha}_{.~~\mu \nu} {\stackrel{def.}{=}}~ U^{\alpha}_{.~(\mu \nu )} + U^{\alpha}_{.~~[\mu \nu]} 
=  \cs{\mu}{\nu}{\alpha}\ + K^{\alpha}_{.~~\mu \nu}   
\end{equation}  
where
\begin{equation}
U^{\alpha}_{.~(\mu \nu)} {\stackrel{def.}{=}}~ \frac{1}{2}(~U^{\alpha}_{.~~\mu \nu } + U^{\alpha}_{.~~\nu \mu }) ,
\end{equation}
\begin{equation}
U^{\alpha}_{.~~[\mu \nu]} {\stackrel{def.}{=}}~ \frac{1}{2}(~U^{\alpha}_{.~\mu \nu } - U^{\alpha}_{.~ \nu \mu})
=  K^{\alpha}_{.~[\mu \nu]} = \frac{1}{2} S^{\alpha}_{.~\mu \nu}   ,
\end{equation}	 
where $S^{\alpha}_{.~\mu \nu}$ is a third order tensor known as the torsion tensor.

  The contravariant metric tensor $g^{\mu \nu}$ is defied such that 
\begin{equation}
g^{\mu \alpha}g_{\nu \alpha} = g^{\alpha \mu}g_{\alpha \nu} = \delta^{\mu}_{\nu}  .
\end{equation}    
The covariant derivatives {(21), (22) and (23)} are connected to the parameter derivatives by the
relations 
\begin{equation}
\frac{\nabla A^\mu}{\nabla \tau^{-}} = A^{\mu}_{-|||~ \alpha}\tilde{J}^\alpha,
\end{equation}
\begin{equation}
\frac{\nabla A^\mu}{\nabla \tau^{0}} = A^{\mu}_{0|||~ \alpha}\tilde{W}^\alpha,
\end{equation}
\begin{equation}
\frac{\nabla A^\mu}{\nabla \tau^{+}} = A^{\mu}_{+|||~ \alpha}\tilde{V}^\alpha , 
\end{equation}
where $\tilde{J}^{\mu}$,$\tilde{W}^{\mu}$ and $\tilde{V}^{\mu}$ are tangents to the paths whose parameters are
$\tau^{-}$,$\tau^{0}$ and $\tau^{+}$ respectively.\\ 
\section*{\bf 4. Quantum Features of Einstein's Geometry}

Applying the Bazanaski approach [4] to the Lagrangian functions
\begin{equation}
\Xi^{-} = g_{\mu \alpha}\tilde{J}^{\mu}\frac{\nabla \Psi^\alpha}{\nabla\tau^{-}} ,
\end{equation}
\begin{equation}
\Xi^{0} = g_{\mu \alpha}\tilde{W}^{\mu}\frac{\nabla \Theta^\alpha}{\nabla\tau^{0}} ,
\end{equation}
\begin{equation}
\Xi^{+} = g_{\mu \alpha}\tilde{V}^{\mu}\frac{\nabla \Phi^\alpha}{\nabla\tau^{+}} , 
\end{equation}
where $\Psi^\alpha, \Theta^\alpha$ and $\Phi^\alpha$ are the deviation vectors, we get the following set path equations respectively: 
\begin{equation}
{\frac{d\tilde{J}^\alpha}{d\tau^{-}}} + \cs{\mu}{\nu}{\alpha}\ \tilde{J}^{\mu}\tilde{J}^{\nu} = -K^{\alpha}_{.~\mu\nu}\tilde{J}^{\mu}\tilde{J}^{\nu} ,
\end{equation}
\begin{equation}
{\frac{d\tilde{W}^{\alpha}}{d\tau^{0}}} + \cs{\mu}{\nu}{\alpha}\ \tilde{W}^{\mu}\tilde{W}^{\nu} =  -\frac{1}{2}
g^{\alpha \sigma}g_{\mu \rho}S^{\rho}_{.~\nu \sigma} \tilde{W}^{\mu}\tilde{W}^{\nu} - K^{\alpha}_{.~\mu\nu}\tilde{W}^{\mu}\tilde{W}^{\nu} ,
\end{equation}
\begin{equation}
{\frac{d\tilde{V}^\alpha}{d\tau^{+}}} + \cs{\mu}{\nu}{\alpha}\ \tilde{V}^{\mu}\tilde{V}^{\nu} =
 - g^{\alpha \sigma}g_{\mu \rho}S^{\rho}_{.~\nu \sigma} \tilde{V}^{\mu}\tilde{V}^{\nu} - K^{\alpha}_{.~\mu\nu}\tilde{V}^{\mu}\tilde{V}^{\nu}.
\end{equation}	 
This set of equations can be written in the following general form: 
\begin{equation}
{\frac{dC^\alpha}{d\tau} +a~ \cs{\mu}{\nu}{\alpha}\ C^{\mu}C^{\nu} =
- b~ g^{\alpha \sigma} g_{\mu \rho} S^{\rho}_{.~\nu \sigma} C^{\mu}C^{\nu} - 
c~ K^{\alpha}_{.~\mu\nu} C^{\mu} C^{\nu}},
\end{equation}
where $a$, $b$ and $c$ are the numerical coefficient of the Christoffel symbol, torsion and K-terms
respectively. Thus, we can construct the following table.
\begin{center}
\newpage
  Table II: Coefficients of The Path Equations in Einstein Non-Symmetric Geometry \\ 
\vspace{0.5cm}                                                                                                                                                                                                                                                                                                                                                                                                                                               
\begin{tabular}{|c|c|c|c|} \hline                                                                                                                                                                                                                                                                                                                                                                                                           
Affine Connection  &Coefficient $a$        &Coefficient $b$     &Coefficient $c$   \\
\hline
& & & \\
${\hat U}^{\alpha}_{.~\mu \nu}$    & 1   & 0  & 1 \\
& & & \\ \hline
& & & \\ 
$U^{\alpha}_{.~(\mu\nu)}$       & 1   & $\frac{1}{2}$  &1  \\
& & & \\ \hline 
& & & \\
$U^{\alpha}_{.~\mu \nu}$ 
&1    & 1      &1  \\
& & & \\ \hline  
 \end{tabular}
\end{center}
\section*{\bf 5. Discussion and Concluding Remarks}  

It is generally accepted that gravity is successfully described in the framework of 
geometrization, while other interactions are well understood in the framework of quantization.
However, the marriage between quantization and geometrization has never been successful, so far.
In particular a theory for quantum gravity is still far beyond the reach of researchers.
The problem may be in the roots of general relativity theory or quantum theory.
Some authors (cf. Ref. 5) summarized the problem in the following question:
Can the technical formalism of quantum theory  
handle the idea that space-time itself might have quantum properties in addition
to those of the metric and other fields that might carry? This question is asked from
the quantum point of view. On the other hand, from the geometric point of view,
we believe that it is better to look for geometries that admit some quantum
features. In other words, before trying to quantize general relativity it is better to 
examine its roots i.e. the Riemannian geometry, or other geometries in which the theory 
might be written. If some quantum features are discovered in a certain geometry,
then the theory should be written in this geometry as a first step before any 
quantization.

 In a trial to explore different paths admitted by the AP-geometry [3],
it is shown that three path equations are admitted including the geodesic one.
The three paths are geodesic equations modified by a torsion term whose numerical coefficient
jumps by a step of one half from an equation to the next as mentioned in the introduction.
It has been tempting to believe that the paths in the AP-geometry have some quantum 
features {\it {i.e." the jumping value of the coefficient of the torsion term"}}. In looking for a physical meaning for the torsion term, it has been shown 
[6], that this term should first be parametrized in order to be consistent with the 
observational basis of general relativity. The parameter has been suggested to include the fine structure
constant. The torsion term is then interpreted as representing a type of interaction
between the background gravitational field, in which the particle moves, and the
quantum spin of this particle.

   Table I shows that in the AP-geometry the discovered quantum features are closely connected to the
torsion of space-time, whatever the symmetric part of the connection is (compare 
the second row and the fifth row in this table). Table II 
shows that the same features are present in the Einstein non-symmetric geometry
and connected also to the torsion term,
although the geometry used in this table is quite different from that used in Table
 1. This shows clearly why these features are not present in case of Riemannian
geometry. Now, we can answer the questions given in the introduction.\\
{\underline{ For the first question:}} In AP-geometry, by using different affine connections, 
we get the same equations obtained in the previous work [3].\\
{\underline {For the second question:}} As we get the same equations, the value
of the numerical step remains unchanged. It is one half as we go from one path 
equation to the next.\\  
{\underline {For the third question:}} Although the resulting path equations of Einstein
non-symmetric geometry are different from those of AP-geometry, the same quantum 
features are present in this geometry. The numerical step of the torsion term
remains the same as in AP-geometry (compare the third columns of Tables I and II).

   Now we can conclude this work by stating that some quantum features of the paths 
of non-symmetric geometries are closely connected to the torsion term (non-symmetric
part of the connection). These features are dependent neither on the type of non-symmetric 
geometry nor on the symmetric part of the connection (compare the third columns of Tables I and II). The extra terms appearing in
the path equations of Einstein non-symmetric geometry (the torsion term and the
K-term) should be parametrized as done in the case of the AP-geometry [6].
Further efforts are needed, from both observational and theoretical points of view, in order to confirm this conclusion. 
\section*{\bf References}  
{1.}  Robertson, H. P. (1932) Annals of Mathematics. Princeton (2), {\bf 33}, 496.\\
${\ } {\ }$Mikhail, F. I.(1952) Ph.D. Thesis, London University\\
${\ }{\ }$M\o ller, C. (1978){ Matematisk-Fysiske Skrifter udgivet af 
 Del Kongelige Danske\\
${\ }{\ }$   Videnskabernes Selskab},A{\bf 39}, 1.\\  
${\ }{\ }$ Hayashi, K. and Shirafuji,T. (1979) Physical Review D{\bf 19}, 3524. \\
{2.}  Einstein, A. (1955) {\it The Meaning of Relativity}, Princeton Univ. Press.\\
${\ }{\ }$ Moffat, J. W. (1995) Journal of Mathematical Physics {\bf 36}, 3722.\\
{3.}  Wanas, M. I. , Melek, M. and  Kahil, M. E. (1995){ Astrophysics  and Space Science},{\bf 228},

 273. \\
{4.}   Bazanski, S. L. (1977){ Annales de l'Instiut Henri Poincar\'e}, Section-A,{\bf 27}, 145 .\\
${\ }{\ }$ Bazanski, S. L. (1989) Journal of Mathematical Physics, {\bf 30}, 1018.\\
{5.}   Isham, C. J. (1995) Proceedings of GR14 ,P.167 editors M. Francviglia et al.,\\
${\ }{\ }$ World Scientific Publisher.\\
{6.}   Wanas, M. I. (1998){ Astrophysics  and Space Science},{\bf 258}, 237. \\

\end{document}